\documentclass[12pt,twoside]{article}
\input epsf
\usepackage{natbib}
\usepackage{color}
\usepackage{epsfig}
\usepackage{pictex}
\usepackage{t1enc}
\usepackage[latin2]{inputenc}

\newcommand{\psfigure}[3]{
        \begin{figure}[h!]
        \vspace{0.2in}
        \begin{center}
        \epsfxsize=#2 \epsfbox{#1.eps}
        \end{center}
        \caption{#3}
        \label{fig:#1}
        \end{figure}
}
\newcommand{\psfigureabv}[5]{
        \begin{figure}[h!]
        \vskip 0.2in
        \begin{center}
	\begin{tabular}{c}
	    \epsfxsize=#4 \epsfbox{#2.eps} \\
		(a) \\
		~ \\
        \epsfxsize=#4 \epsfbox{#3.eps} \\
		(b) \\
	\end{tabular}
        \end{center}
        \caption{#5}
        \label{fig:#1}
        \end{figure}
}
\newcommand{\psfigureabhs}[7]{
        \begin{figure}[h!]
        \vskip 0.2in
        \begin{center}
	\begin{tabular}{cc}
	    \begin{tabular}{c}
	    	~\\[#6]
	    	\epsfxsize=#3 \epsfbox{#2.eps} \\
	    \end{tabular} &
        \epsfxsize=#5 \epsfbox{#4.eps} \\
		(a) & (b) \\
	\end{tabular}
        \end{center}
        \caption{#7}
        \label{fig:#1}
        \end{figure}
}

\newcommand{\psfigureabw}[5]{
        \begin{figure}[h!]
        \vskip 0.2in
        \begin{center}
	\begin{tabular}{ccc}
	        \epsfxsize=#4 \epsfbox{#2.eps} &
	        &
        	\epsfxsize=#4 \epsfbox{#3.eps} \\
		(a) & & (b) \\
	\end{tabular}
        \end{center}
        \caption{#5}
        \label{fig:#1}
        \end{figure}
}

\begin{document}
\vspace{0.1in}
\begin{center}
\Large
	An Algorithm for Transforming \\
	Color Images into Tactile Graphics
\vspace{0.2in}

\small
	Technical Report IITiS-20040408-1-1 \\
	April 8th, 2004 \\

\vspace{0.2in}

	Artur Rataj \verb|<arataj@iitis.gliwice.pl>| \\
	Institute of Theoretical and Applied Computer Science \\
	of the Polish Academy of Sciences, \\
	Ba\l tycka 5, 44--100 Gliwice, Poland
\end{center}

\vspace{0.1in}

\begin{quote}
\small
	\textbf{Abstract.}
	This paper presents an algorithm that transforms color visual images,
	like photographs or paintings, into tactile graphics. In the algorithm,
	the edges of objects are detected and colors of the objects are
	estimated.  Then, the edges and the colors are encoded into lines and
	textures in the output tactile image. Design of the method is
	substantiated by various qualities of haptic recognizing of images.
	Also, means of presentation of the tactile images in printouts are
	discussed. Example translated images are shown.
	
\end{quote}

\vspace{0.0in}

\section{Introduction}
	Tactile graphics \citep{hinton1966tactile, edman1992tactile}
	is a way of obtaining information about images by blind
	people. However, visual images can use features like colors and
	textures, or edges with widely varying strength, that can be difficult
	to translate into tactile form. One of the reasons of it can be that the
	details in a tactile image may need to be larger than the size of
	respective details in the visual image, a blind person may sense only
	parts of the tactile image at a time, and special features may be needed
	in the tactile image to encode colors and help guide the fingers.
	
	The translation of images into the tactile form is often performed by
	skilled professionals, especially in the case of transforming
	technical materials like maps or diagrams. However, a human
	translator may not always be available or translation by a human
	may be not fast enough. In these cases, a computer doing automated
	transformation of visual images into tactile images may be helpful.
	Also, such a computer might also help a human translator, by
	performing some tasks that would form a base to be fine--tuned or extended
	by a human. There are many methods of converting images into
	tactile graphics \citep{hinton1966tactile, edman1992tactile,
	eriksson1988tactile, gill1999tactile, eriksson1999tactile}.
	
	In this paper, an automated method of transforming color images like
	photographs or paintings, in contrast to specialized images like maps,
	is proposed. In the method, edges and colors of objects in the input
	image are translated into appropriate features in the output tactile
	image. The edges are translated into varying thickness lines, to help
	distinguishing objects in the tactile image. A way of
	translating colors into textures is used that is the same for all
	of the processed images. This is because a color is not
	only a visual feature -- it has also a deep semantic meaning. A blind
	person can, for example by talking or reading books, meet the notion of
	color, of course, and get acquainted with facts like that the sun is yellow,
	sky is blue and grass is green, like a sighted person. In effect,
	conveying the information about colors may be one of the most important
	methods of helping the blind person in recognizing an image, especially
	in the case of non--technical images like photographs.
	
	The paper also describes a simple method of presentation of the resulting
	tactile image on a printout.

	Some examples of translation of color images using the described method are
	shown.
	
\section{The algorithm}
\label{sec:algorithm}
	In this section, first the edge encoding is described, and then
	classifying and encoding of colors is presented.

\subsection{Processing edges}	
\label{sec:edges}
\begin{figure}[h!]
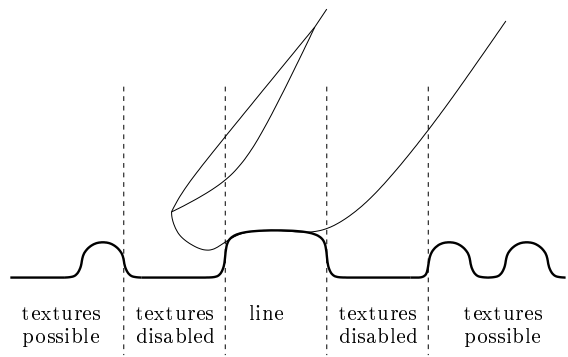

        \vspace{0.2in}
        \begin{center}
	\input edge.pstex_t
        \end{center}
        \caption{A schematic example of a line sensed by a finger.}
        \label{fig:edge}
        \end{figure}

	A special edge detector, developed by the author,
	is used in the presented method. The detector
	introduces a relatively low level of artifacts into its output image
	of edges and provides information about the strength of the edges.
	To not clutter the tactile image, only edges above a given strength are
	encoded. An edge is encoded with a line whose thickness increases
	with the local strength of the edge. The thinnest lines are
	thick enough to be sensed relatively
	well. The thickest lines are thin enough to guide a finger.
	Around a line, there is a small always non--textured gap so to
	ease sensing of the line and guiding of fingers, like it is show in
	Fig.~\ref{fig:edge}.
	
	The thickness of the lines reaches a maximum value with the strength of
	the represented edges relatively fast, so that most of the edges in
	an image are often of the maximum thickness. This may ease finding edges,
	because of their repeated thickness. On the other hand, fine lines,
	to represent some less visible details, are still available and an
	edge that slowly and gradually gains its strength in the color image
	may be represented by a line in the tactile image that gradually
	increases its thickness, instead of beginning abruptly, what could
	be confusing.
		
\subsection{Processing colors}
\label{sec:colors}
	Colors in the presented method are encoded with different textures.

	The initial step in encoding colors is their blurring so that very dense
	patterns in the input color image are approximated with locally
	averaged colors. 
	This is performed to simplify the estimation of colors of objects,
	because it substitutes the averaging quality of human vision. For example, a
	lawn consisting of green, black and yellow details may generally be
	perceived as having an average color of dark, yellowish green. In the
	case of haptic sensing, a human could need to estimate the average color
	by trying to associate small, differently textured regions,
	what could be slow and cumbersome.
	Also, haptic sensing may be less precise in recognizing
	small mixed areas of different textures. The blurring of colors is performed
	after the edge detection, so that lines may still `sharply' represent
	borders between objects.
	
	The resulting blurred colors are quantized into discrete groups. There
	are four groups representing saturated colors: blue, green, yellow and
	red, and a fifth group of black, grays and slightly saturated colors.
	In the quantization process, a color representation is changed from
	the RGB color space \citep{sharma97digital} to the HSV color
	space\citep{schwarz1987colormodels}. Then, there are some
	correction of the hue component on basis of the saturation and value
	components -- it is taken into account that a color having the hue component
	of yellow may become green when its value component decreases, and similarly
	darkening a color with the hue component of orange may make the
	color yellowish.

	Human vision has specialized mechanisms of texture
	classification and segmentation\citep{bergen91computational,
	malik1990texture} yet the situation is different with
	haptic perception. Textures easy and fast to classify using one of the
	two discussed ways may be not the easiest one to perceive by using
	another of the ways. Because in coding colors as textures, tactile image
	may contain large textured regions, in the design of the proposed
	algorithm an emphasis was put on fast recognition of different textures. 
	Another problem addressed was such a definition of the textures that
	they could be relatively
	easy to distinguish, in typical images, from the lines that encode
	edges.

	The proposed textures are designed to be easily distinguished by moving
	fingers over the tactile image. Thus, a human moving fingers to locate
	the edges may also relatively fast classify the textures without halting
	or slowing the fingers. Also, a way of helping locating fast the lines
	by varying the density of features in the textures was developed.
	
	Accordingly to the quantization of colors, there are five classes of the
	textures.  The classes representing saturated colors have the same
	pattern each, but the pattern is oriented at different angles in the
	case of each these classes. The pattern has strong directional quality,
	so that the class of texture may be easily determined by sliding fingers
	over the tactile pattern. The fifth class, of black, grays and
	lowly saturated colors, has weak directional
	properties, what makes the class, by contrast, easily distinguishable
	from the four other classes.
	
	Within each of the five classes, the `boldness' of the
	features is inversely proportional to the brightness of the represented
	color, so that there is a smooth transition to white in the case of each
	class, what reduces abrupt changes that might be sensed as an artifact.
	
	Samples of the discussed textures are shown in Fig.~\ref{fig:textures}.
	\psfigureabv{textures}
		{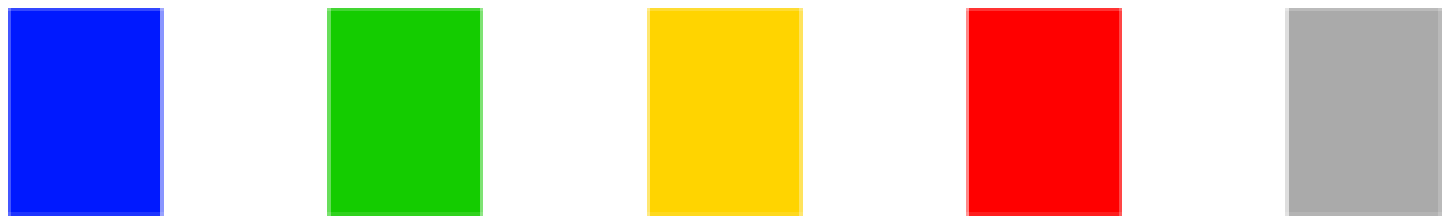}
		{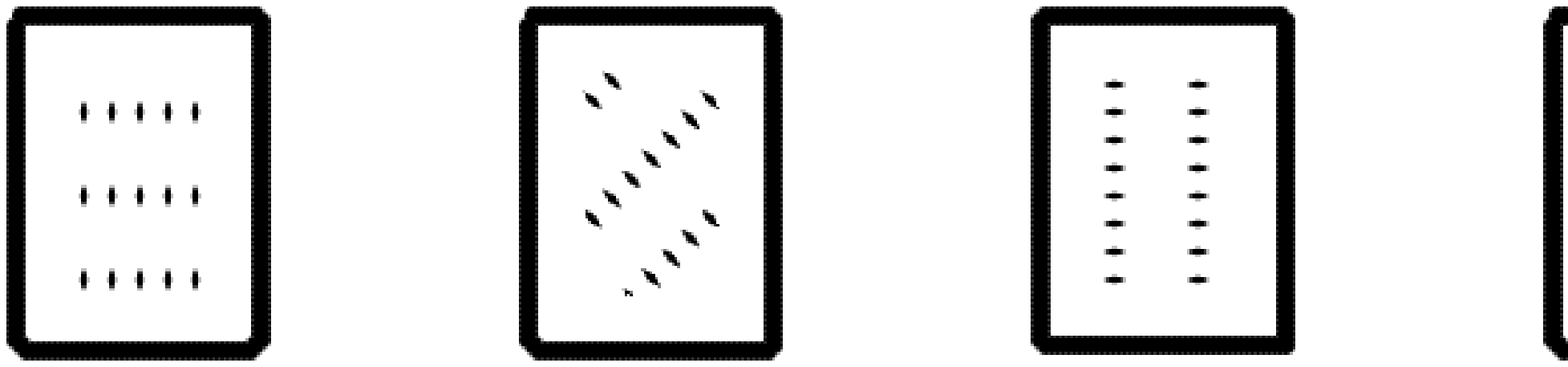}
		{3in}
		{An input image (a) with five bars: blue, green, yellow, red
		and gray, on white background, and the resulting (b) tactile image.}
	It can be seen that the textures representing saturated colors consist
	of small rhombus--like features, and the fifth class, representing colors
	not saturated or slightly saturated, consists of small dot--like
	features. This way, the textures, by their `bumpy' quality, are more
	distinguishable from the lines that represent edges.  The shapes of the
	features further enhance the directional quality of the
	textures representing saturated colors.
	
	The reasons for assigning such textures as can be seen in
	Fig.~\ref{fig:textures}, apart from the reasons already mentioned, were
	as follows:
	\begin{itemize}
		\item
			The less saturated colors are often perceived as more neutral,
			thus the weak directional properties of the class representing
			the grays or slightly saturated colors, and highly directional
			properties of the saturated colors. The directionality of
			the classes representing saturated colors matches also
			the need of distinguishing between these classes.
			
		\item
			The gradual change of orientation of the textures in the
			saturated color classes matches the order of the colors in the
			color wheel.
		
		\item
			Blue can be associated with horizontal directional, because
			water basins like rivers or lakes often have their surface
			horizontal or very close to the horizontal.
		
		\item
			Most people are right--handed, what make it likely that they
			will more readily use their right hand to detect the textures
			a tactile image.
			Moving fingers of the right hand over the tactile image that
			does not require larger movements of the wrist, 
			that is require a relatively small amount of energy, is smoother in
			the case of the `green' class of textures in compare to
			the `red' class, as it is shown in
			Fig.~\ref{fig:fingers-movement}. The smoothness of texture, in turn,
			can make it easier to detect lines representing the edges of objects.
			It may have two advantages. First, green
			generally may be more common in many images than red. Second,
			green may be perceived as a more `calm' color than red, what
			matches well with the discussed smoothness of moving fingers.
			
        \begin{figure}[h!]
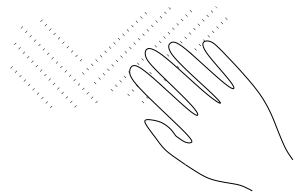

        \vspace{0.2in}
        \begin{center}
	\input fingers-movement.pstex_t
        \end{center}
        \caption{An example of a likely way of placing a right hand
				 by a right--handed person over a tactile image
				 with textures representing red and green.}
        \label{fig:fingers-movement}
        \end{figure}

	\end{itemize}
	
	Within each class of textures, not only the boldness, but also the
	density of the features may vary. The density of a texture expresses
	local mean density of lines in the tactile image. In the relatively empty
	regions the textures are sparse, and near lines the textures are dense. 
	That makes finding lines in the tactile image faster, because the
	regions near to edges are emphasized by the textures, and the regions far from edges may
	be sensed faster because of the low density of features.  The
	greater density of the color encoding textures near to the lines
	increases also the precision of representing colors, what may be
	important because it may be likely that near to the lines there are
	small regions limited by these lines. In regions very near to the lines,
	the textures are disabled, of course, as it was discussed in
	Sec.\ref{sec:edges}. An example of varying density of textures is shown
	in Fig.~\ref{fig:textures-density}.
	\psfigure{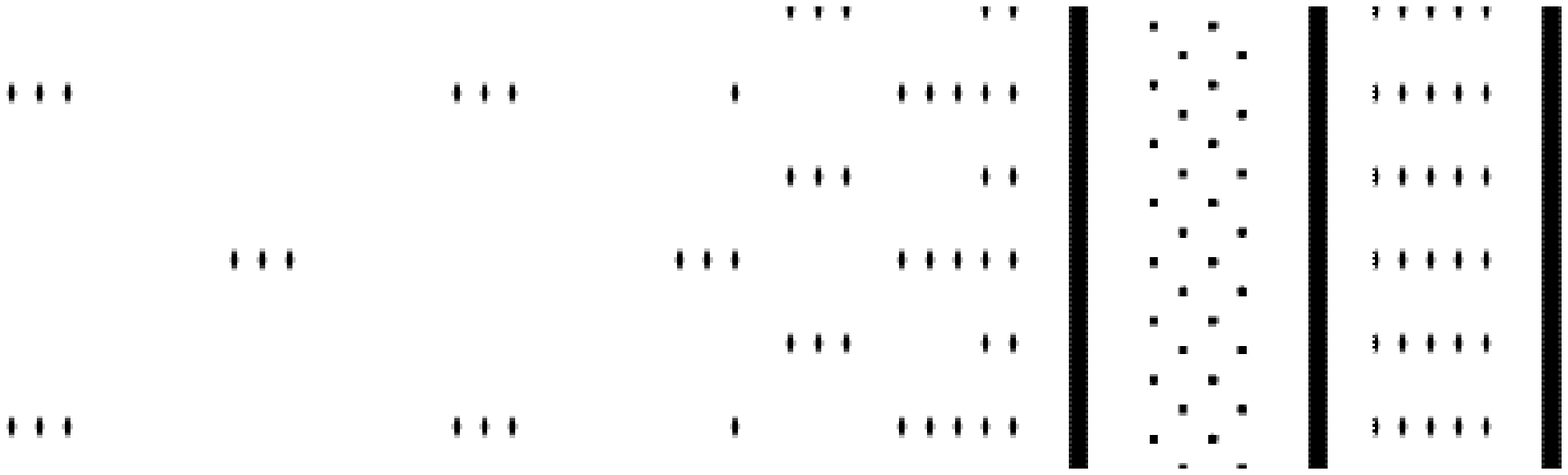}{3in}
		{Example tactile image with textures having varying density.}
	The local mean densities are computed as follows. A quantized image of
	edges is created, that is an image in which every edge with a strength
	above a given threshold, the same as the threshold discussed
	in Sec.~\ref{sec:edges}, is marked with 1, and the rest of the image is
	marked with 0. Then the image is blurred. The resulting values in the
	image, in the range from 0 to 1, determine the discussed densities
	of textures.
	
	The number of classes of the textures is limited to five to increase the
	speed of recognizing the classes, what may be important because an image
	may contain large textured regions. Some experiments shown that more
	classes may make the task substantially slower. On the other hand, it
	may be advantageous to quantize colors to more than five groups. 
	Perhaps a way of maintaining the speed of recognizing, while at the same
	time representing colors by more classes of textures, could be done by
	dividing a `main' texture class into subclasses, that differ by some
	`secondary' features, in opposite to the `primary' feature, that
	determines a whole `main' class. The secondary features would need a
	closer inspection to recognize, but on the other hand these features
	would not interfere with the recognition of the primary features. 
	This way, determining the the `main' class would still be fast, but
	there would still be a possibility of determining a `finer' subclass.
	
\begin{figure}[h!]
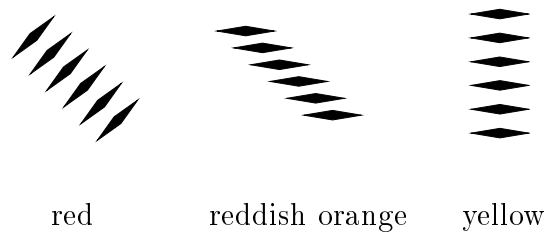

        \vspace{0.2in}
        \begin{center}
	\input second-order-features.pstex_t
        \end{center}
        \caption{An example of representing `typical' and `fringe' colors by textures.}
        \label{fig:second-order-features}
        \end{figure}

	One of propositions of introducing the secondary features is to
	divide each of the four classes representing the saturated colors into
	three subclasses. The `center' subclass would represent a `typical' blue,
	green, yellow or red, and the `fringe' classes would represent
	`transitional' colors. The differentiating secondary feature would be
	the orientation of the rhombuses, i.~e.~reddish orange would be
	represented by a texture in the `main' class of the red color, yet the
	rhombuses of the texture would be oriented as in the case of a `typical'
	yellow. The example is illustrated in Fig.~\ref{fig:second-order-features}.
	In the current version of the algorithm, the `fringe' classes are not
	used and it will be tested in experiments whether the optional introduction
	could be advantageous in some cases.
	
\subsection{Presenting an image on a printout}
\label{sec:presenting}
	The images are printed withing a textured frame. There is a gap between
	the frame and the image, so that the frame does not interfere with the
	image contents. The texture of the frame is like the class of textures
	representing grays, yet much finer, to be easily distinguishable. The
	very weak directional properties provide neutrality of the frame. Below
	the main tactile image with lines and textures, there is a smaller
	tactile image
	with lines only, also in a frame. The smaller image can be used for
	rough yet fast determining of the image contents, what may be helpful
	when recognizing the larger, textured image. Between the frames of the
	two images there is a substantial gap, to ease fast locating of the
	two tactile images.

\section{Tests}
\label{sec:tests}
	Let us first test the drawing shown in Fig.~\ref{fig:landscape}(a).
	The output tactile form of the image, transformed by the discussed
	algorithm and presented in the form for printing,
	is shown in Fig.~\ref{fig:landscape}(b). 
	\psfigureabhs{landscape}
		{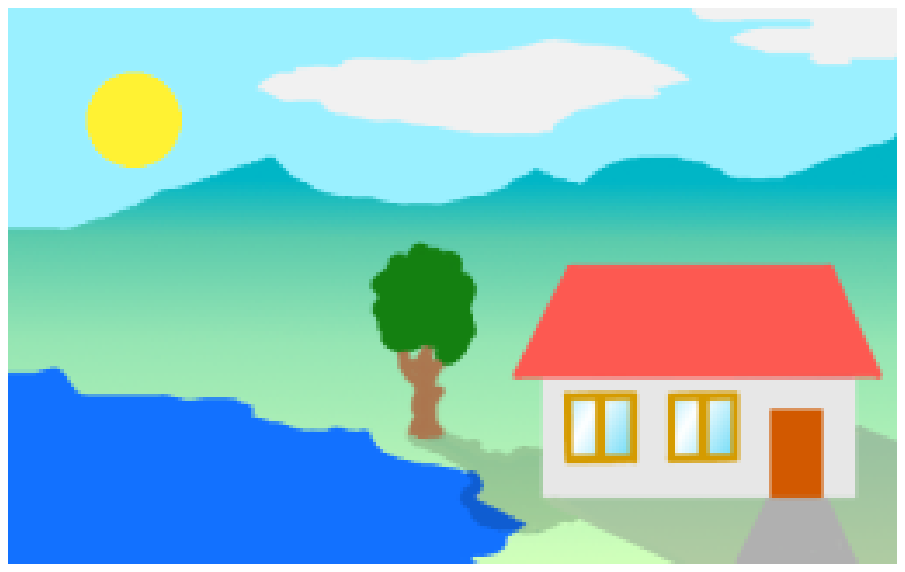}{1.8in}
		{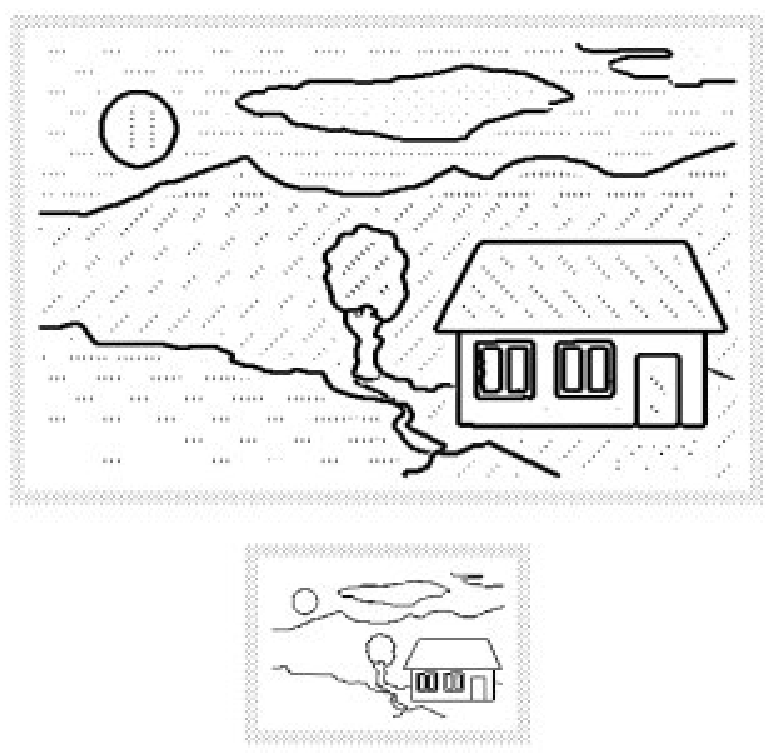}{2in}
		{-2in}
		{An example translation of a color image (a) into a tactile
		graphics printout (b).}
	It is obvious that the ability of telling the colors of object
	in the image like the
	pond, the sun or the grass, among other, may help in recognizing the
	objects. 

	Let us test the translation of a sample photograph, where the varying
	thickness of lines is more clearly visible.
	An example such a translation is shown in Fig.~\ref{fig:clock-tower}.
	\psfigureabw{clock-tower}
		{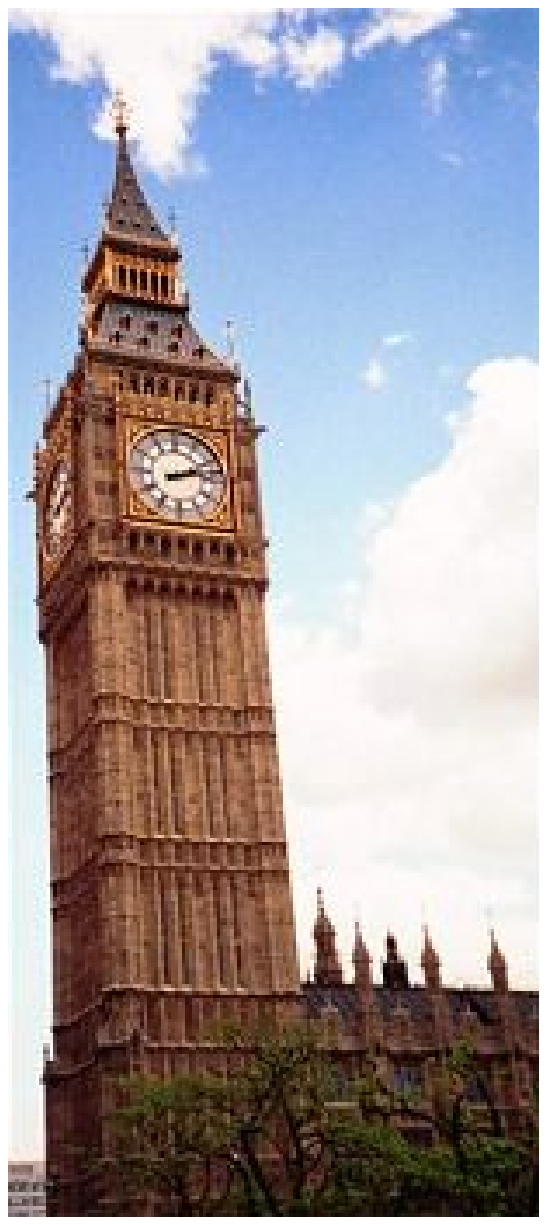}
		{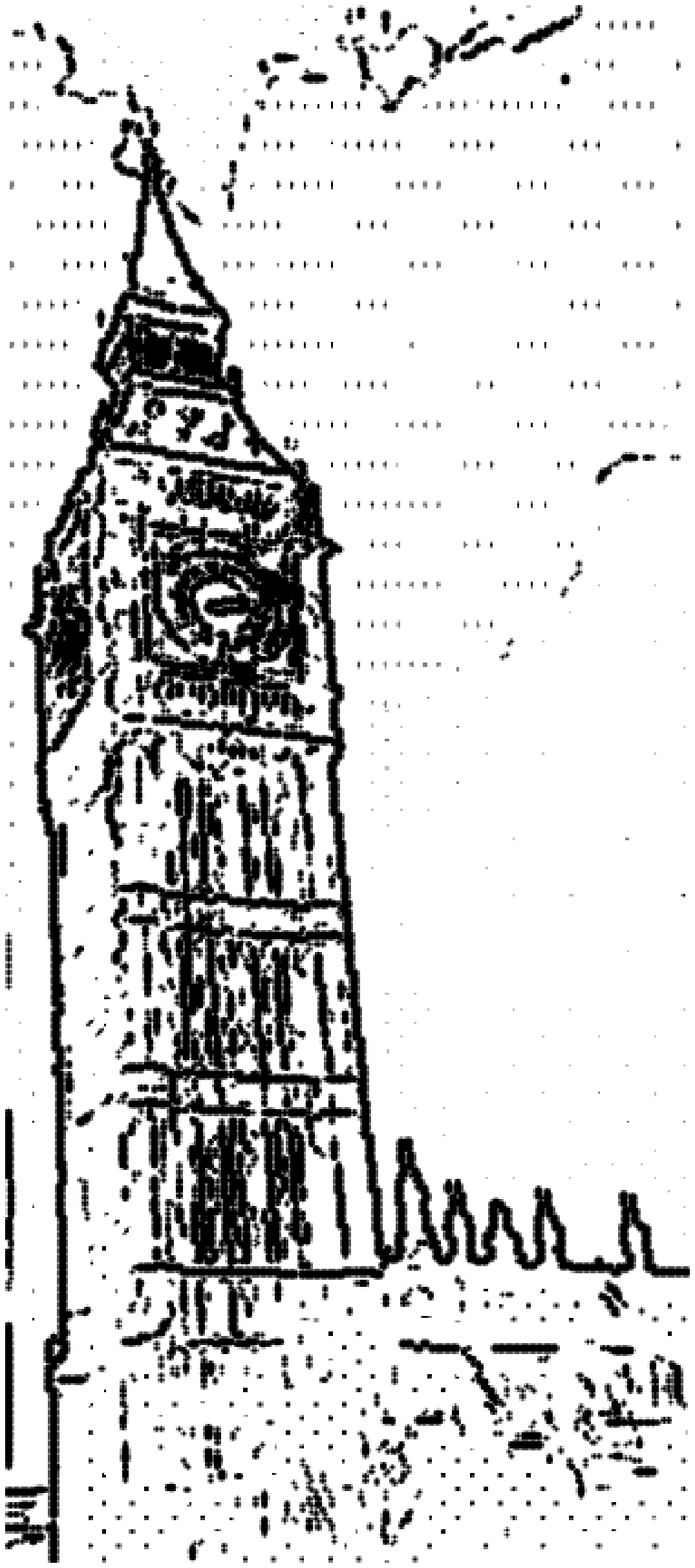}
		{1.4in}
		{Example conversion of a color photograph (a) into
		a tactile image (b).}
	It can be seen in the example that while several features may be easily
	detected, some fine--tuning of the edges by a human could improve the
	tactile image.

\section{Conclusions}
\label{sec:conclusions}
	A human translator may in some cases create higher quality images than
	the presented automatic method would do, because a human can understand
	an image on a semantic level. For example, the edge detector used in the
	method is color gradient--sensitive, while a human could mark edges in
	the tactile image by basing on the semantic meaning of the edges. On the
	other hand, an automatic translation of color graphics into tactile form
	may speed up often long--lasting tasks like tracing object boundaries or
	marking objects with textures, and very precisely draw the textures
	having variable density and variable boldness. The intermediate solution
	is to use a computer as a help to a human translator. For example, a
	human translator could mark the edges or alter, in a graphics
	application, the image of edges provided by the algorithm, and then the
	algorithm could complement the image with textures. Summarizing, the
	preferred method may depend on things like image complexity,
	availability of a human translator, the required quality or the required
	speed of the translation.
	
	The presented method is currently consulted with the blind people.
	Various parameters may change, like the thickness of lines, the width of
	the regions with disabled textures around the edges, or the scale of the
	textures. Also the possibility of introducing of the discussed `fringe'
	colors is consulted.
	
\bibliography{nn}
\bibliographystyle{apalike}

\end{document}